\documentclass[aps,prl,reprint,superscriptaddress,preprintnumbers,x11names]{revtex4-2}
\pdfoutput=1

\usepackage{amsfonts, amssymb, amsmath}
\usepackage{graphicx}
\usepackage{mathrsfs}
\usepackage{hyperref}
\usepackage{booktabs}
\usepackage{slashed}

\usepackage[utf8]{inputenc}
\usepackage[T1]{fontenc}
\usepackage{lmodern}
\usepackage{microtype}

\usepackage{xcolor}

\usepackage{graphicx}
\usepackage[percent]{overpic}
\usepackage{caption}
\usepackage{subcaption}

\usepackage{hyperref}
\hypersetup{
	colorlinks=true,
	linkcolor=Blue4,
	citecolor=Red4,
	urlcolor=Green4,
	linktoc=page
}

\usepackage[paperwidth=210mm,paperheight=297mm,centering,hmargin=1.5cm,vmargin=2.5cm]{geometry}
\newcommand{\rmi}{i}
\newcommand{\rme}{\mathrm{e}}
\newcommand{\rmd}{\mathrm{d}}
\newcommand{\dd}{\mathrm{d}}

\newcommand{\f}{\frac}

\newcommand{\Z}{\mathbf{Z}}

\begin{document}

\title{Localisation of the M2-brane}

\date{\today}

\author{Fri{\dh}rik Freyr Gautason}

\affiliation{STAG Research Centre $\&$ Mathematical Sciences, University of Southampton Highfield, Southampton SO17 1BJ, U.K.}

\author{Jesse van Muiden}
\affiliation{Theoretical Physics Group, Blackett Laboratory Imperial College, London SW7 2AZ, U.K.}

\begin{abstract}
\noindent 
We study quantum M2-branes in holographic backgrounds and show that their moduli spaces of zero-modes are localised according to an R-symmetry Killing vector. We discuss the relation with recent results in the context of equivariant localisation in gauged supergravity and argue its origin within M-theory path integrals expanded in saddle points over M2-branes. We argue that the M2-brane partition function, including its non-perturbative corrections, should be compared to the field theory grand canonical partition function. Our results extend recent observations in the context of the giant graviton expansion of superconformal indices to generic supersymmetric AdS boundary conditions. As a byproduct we predict non-perturbative corrections to a variety of supersymmetric observables of the ABJM theory.
\end{abstract}


\maketitle

%%%%%%%%%%%%%%%%%%%%%%%
\section{Introduction}\label{sec:intro}
%%%%%%%%%%%%%%%%%%%%%%%
The strong coupling limit of string and M-theory has largely remained an enigma due to the lack of tools that allows for the explicit computation of physical observables. This parallels the early history of quantum field theories, although in recent decades there has been major progress through the development of modern tools such as integrability, bootstrap, and localisation to name a few. The latter has proved invaluable in the study of non-perturbative dynamics of supersymmetric quantum field theories, and allowed for the exact determination of partition functions, correlation functions, and expectation values of line operators \cite{Pestun:2016zxk}.

The successes in supersymmetric quantum field theories can to some extent be translated to guiding principles in string and M-theory   
through top-down holographic dualities. A vital consistency check for the quantisation of strings and branes 
is the comparison of their path integrals in a saddle point approximation to the field theory counterparts computed using e.g. supersymmetric localisation. One example where this interplay has proven particularly fruitful is the duality between M-theory on AdS$_4 \times S^7/\mathbf{Z}_k$  spacetime and the $\text{U}(N)_k \times \text{U}(N)_{-k}$ ABJM theory, with  CS-level $k$ \cite{Aharony:2008ug}. After a map of parameters, the holographic duality states that
\begin{equation}\label{Eq: QG saddle expansion}
	\log Z_{\text{QFT}} = \log \mathcal Z_{\text{M-theory}	} \approx\sum_{\text{saddles}} Z_{\text{1-loop}} \rme^{-S_{\text{cl}}}\,,
\end{equation}
where we have expressed the gravitational path integral as an infinite sum over saddle points. 
Following the intuition from the off-shell background field approach to string theory \cite{Fradkin:1984pq,Fradkin:1985fq,Fradkin:1985ys}, the saddles are written in terms of M-branes with an on-shell action $S_\text{cl}$ and the quantum effective action $Z_{\text{1-loop}}$ is truncated at one loop \footnote{We emphasise that in this background field approach the target space is taken to be off-shell, this approach is not to be confused with the off-shell methods of string field theory.}. The leading saddle is given by the degenerate (point-like) brane with a vanishing classical action and bosonic zero-modes represented by the brane position. 
The one- and higher loop partition functions integrated over the zero-modes are expected to give rise to the supergravity action and its higher derivative corrections. Subleading saddles contribute non-perturbatively and represent brane instanton corrections. Although it is not clear that such an interpretation is valid in M-theory, there is computational evidence supporting it. For example, for 1/2-BPS $S^2 \times S^1$ boundary conditions on AdS$_4$,  the exponentiation of \eqref{Eq: QG saddle expansion} reproduces the giant graviton expansion of the 1/2-BPS index \cite{Arai:2020uwd,Gaiotto:2021xce,Beccaria:2023cuo}. Another example is provided by the ABJM partition function on $S^3$ which was recently shown to receive non-perturbative corrections that can be computed by quantizing an M2-brane instanton \cite{Drukker:2010nc,Hatsuda:2012dt,Gautason:2023igo,Beccaria:2023ujc}. 

In this letter we argue that brane expansions such as the one in \eqref{Eq: QG saddle expansion} are applicable to much more general supersymmetric setups 
\footnote{In the case of giant graviton expansions one can understand the brane expansion to arise from trace relations in the dual quantum field theory. For generic boundary conditions there is, however, no radial quantisation and as such one would have to understand the field theory origin of such an expansion from a Lagrangian point of view, similar to what was suggested in \cite{Lee:2024hef}.}. 
To showcase this we consider eleven-dimensional backgrounds that arise as uplifts of solutions to minimal gauged supergravity in four dimensions. 
We study the M2-brane partition function on these backgrounds in the saddle point approximation and show that supersymmetry implies that the branes are localised to the fixed points of a Killing vector associated with the background Killing spinor. 

Let us start with the leading degenerate saddle, i.e. the on-shell supergravity action itself.
In \cite{BenettiGenolini:2019jdz,BenettiGenolini:2023kxp} it was argued that the four-dimensional supergravity action allows for an equivariant localisation, reducing its evaluation to a sum over fixed points of the Killing vector. As we  discuss in this letter, this localisation happens in the zero-mode sector of the degenerate M2-brane and constitutes the zeroth-order check of supersymmetric localisation of M2-branes. We show that the same localisation procedure also applies to the moduli space of zero-modes in the subleading saddles given by non-degenerate M2-branes. 

Having localised the non-perturbative M2-branes, we quantise them and compute their one-loop partition functions. Comparing to the dual $S^3$ partition function we find that the one-loop result agrees with the non-perturbative correction in the grand canonical ensemble. We argue that this one-loop exactness is a consequence of supersymmetric localisation on the M2-brane.

This letter aims to elucidate the existence of equivariant and supersymmetric localisation in the context of M-theory path integrals, and its implications for non-perturbative M2-branes. Computational details are omitted here and will be presented in \cite{futurepaper}.

%%%%%%%%%%%%%%%%%%%%%
\section{Background}\label{Sec: Mtheory background}
%%%%%%%%%%%%%%%%%%%%%
We study M-theory backgrounds that can be consistently truncated down to four-dimensional $\mathcal N=2$ gauged supergravity. The dynamics of the latter is controlled by the standard Einstein-Maxwell action with a cosmological constant \footnote{Throughout we will use the four-dimensional indices $\mu,\nu=1,\dots,4$, the seven-dimensional indices $\alpha,\beta=5,\dots,11$, the eleven-dimensional indices $M,N=1,\dots,11$, and the M2-brane worldvolume indices $a,b = 1 ,\ldots, 3$. Explicit numbered indices (appearing on gamma-matrices and frames) are flat but when there is an ambiguity flat indices are hatted.}:
\begin{equation}\label{4daction}
	S = -\frac{1}{16\pi G_{N}^{(4)}} \int \star_4 \bigg(R_4 + 6 - \frac14 F_{\mu\nu}F^{\mu\nu}\bigg)\,,
\end{equation}
where $F = \rmd A$, and $R_4$ is the Ricci scalar and $\star_4$ denotes the Hodge star operator. The theory also contains a gravitino $\psi_\mu$ whose supersymmetry variation determines the preserved supersymmetry in the background
\begin{equation}
	\delta_\eta \psi_\mu  =  \Big(\nabla_\mu - \rmi \frac{A_\mu}{2} + \frac12 \gamma_\mu   + \frac{\rmi}{4} \slashed{F} \gamma_\mu\Big)\eta = 0\,.
\end{equation}
Here $\nabla_\mu$ denotes the spin-covariant derivative, $\gamma_\mu$ denotes the curved four-dimensional gamma-matrices, $\slashed{F} = \frac12 \gamma^{\mu\nu}F_{\mu\nu}$, and $\eta$ is the supersymmetry parameter. 

Any solution to the equation of motion of the four-dimensional theory can be uplifted to a solution of eleven-dimensional supergravity as follows \cite{Gauntlett:2007ma}
\begin{equation}\label{Eq: 11d metric and three-form}
\begin{aligned}
	\rmd s_{11}^2 = &\, L^2\Big(\rmd s_{4}^2 + 4 \rmd s_{\text{KE}_6}^2 +  (\rmd y + 2\sigma +\tfrac12 A)^2\Big)\,,\\
	G_4 =&\, \rmi L^3(3 \text{vol}_4 + 2 J \wedge \star_4 F)\,,
\end{aligned}
\end{equation}
where $\rmd s_{\text{KE}_6}^2$ is a six-dimensional K\"ahler-Einstein metric with K\"ahler form $\rmd \sigma = 2 J$. In this paper we focus on the holographic dual to the ABJM theory in which case the seven transverse directions combine into a metric on $S^7/\Z_k$ deformed by the four-dimensional gauge field $A$.
More precisely, we choose the following frames on the internal space 
\begin{equation}
\begin{aligned}
	e^{ 5} &= 2L\rmd\theta \,,\\
	e^{ 6} &= L\sin\theta\,\rmd \theta_1 \,,\quad e^{ 7} = L \sin\theta\, \sin\theta_1 \,\rmd\phi_1 \,, \\
	e^{{8}} &= L\cos\theta\,\rmd \theta_2 \,,\quad e^{{9}} = L\cos\theta\, \sin\theta_2 \,\rmd\phi_2 \,,\\
	e^{{10}} &=L \sin\theta\cos\,\theta \,(2\dd\varphi + \cos\theta_1\,\dd\phi_1 - \cos\theta_2\,\dd\phi_2) \,,\\
	e^{{11}} &= L \,\big(\rmd y +2\sigma + \tfrac12 A\big) \,.
\end{aligned}
\end{equation}
where 
\begin{equation}
	2\sigma = \sin^2\theta\,\cos\theta_1\,\dd\phi_1+\cos^2\theta\, \cos\theta_2\,\dd\phi_2-\cos2\theta\,\dd\varphi\,.
\end{equation}
The coordinate ranges are $0\le\theta\le \pi/2$, $0\le\theta_{1,2}\le\pi$, $0\le y,\phi_{1,2}\le 2\pi$, and $0\le\varphi\le4\pi/k$ where the latter implements the $\Z_k$ quotient. It is important to note that we do not quotient the $y$-direction since for non-trivial $A$, we preserve four-dimensional ${\cal N}=2$ supersymmetry and translations along $y$ acts as the R-symmetry. This will become apparent when we write down the Killing spinor associated with the solutions below. %The solution  preserves $\uu(2)^2$ isometry excluding the possible isometries preserved by the four-dimensional metric.

If the four-dimensional solution preserves supersymmetry, then the eleven-dimensional solution preserves (at least) the same amount of supersymmetry. This can be explicitly demonstrated by constructing an eleven-dimensional Killing spinor $\epsilon$ for which the supersymmetry variation of the eleven-dimensional gravitino vanishes: 
\begin{equation}
\delta_\epsilon\Psi_M = \Big(\nabla_M + \f1{24}\Gamma_M\slashed{G}_4- \f18\slashed{G}_4\Gamma_M\Big)\epsilon = 0\,.
\end{equation}
The Killing spinor $\epsilon$ can be identified using similar techniques as in \cite{Larios:2019lxq}. Assuming that it is a product of the conformal Killing spinor on (round) $S^7$ and a four-dimensional spinor, we can reduce the vanishing of the supersymmetry variation along the seven internal coordinates $\delta_\epsilon\Psi_\alpha =0$ to an algebraic condition on the seven-spinor
\footnote{The eleven-dimensional $\Gamma$-matrices are given as a direct product:
$\Gamma_{\hat{\mu}} = (i \gamma_{(4)} \gamma_{\hat{\mu}}) \otimes 1$, and  $\Gamma_{\hat{\alpha}} = \gamma_{(4)} \otimes \gamma_{\hat{\alpha}}$.}. 
This constraint reduces the number of independent components in the seven-spinor from 8 to 2 which can be expressed as
\begin{equation}\label{the7dspinor}
	\chi = \rme^{\rmi y \gamma_{{11}}} \f12\big(1 - \rmi \gamma_{8\,9\,{11}}\big)\f12\big(1-\rmi\gamma_{5\,{10}\,{11}}\big)\chi_0\,,
\end{equation}
where $\chi_0$ is a constant seven-dimensional spinor. This can be further split into positive and negative frequency modes along $y$ using $\gamma_{{11}}$
\begin{equation}
	\chi^\pm \equiv \f12 (1\pm \gamma_{{11}} ) \chi\,.
\end{equation}
The eleven-dimensional spinor can now be written as
\begin{equation}\label{11dspinor}
	\epsilon = \eta \otimes \chi^+ + \eta^c\otimes \chi^-\,,
\end{equation}
where $\eta$ is the four-dimensional Killing spinor satisfying $\delta_\eta\psi_\mu=0$ and $\eta^c$ is its charge conjugate.

We should note that for some particularly symmetric four-dimensional backgrounds, there are more solution to the elven-dimensional BPS equation $\delta_\epsilon\Psi_M=0$. Indeed, the round AdS$_4$ with $F=0$ preserves all 32 supercharges available for $k=1,2$ and 24 supercharges if $k>2$ in eleven-dimensional supergravity. The particular supercharges we have constructed play a special role however, they are the holographic duals to the supercharges used for supersymmetric localisation on the field theory side. Indeed, we will see that it plays a similar role on the gravity side, where it localises the zero-mode sector of instantonic supersymmetric M2-branes.

%%%%%%%%%%%%%%%%%%%%%
\section{Localisation in supergravity}\label{Sec: localisation}
%%%%%%%%%%%%%%%%%%%%%
In \cite{BenettiGenolini:2019jdz} the on-shell action for solutions of minimal gauged supergravity was shown to exhibit a structure reminiscent of supersymmetric localisation in  quantum field theory. This was further elucidated using the Berline-Vergne-Atiyah-Bott (BVAB) fixed point theorem in \cite{BenettiGenolini:2023kxp}.
The argument goes as follows: Any supersymmetric solution of the four-dimensional theory \eqref{4daction} is equipped with a Killing spinor $\eta$ which can be squared to construct a Killing vector $\xi$. From this Killing vector one defines an equivariant exterior derivative $\dd_\xi \equiv \dd -\xi \lrcorner$. A particular set of BPS observables, which includes the on-shell action, can now be reformulated as an integral of an equivariantly closed polyform. 
As a consequence of the BVAB theorem this integral reduces to a sum over the fixed points of $\xi$ \footnote{We are not being careful with boundary terms in this discussion and refer to \cite{BenettiGenolini:2019jdz,BenettiGenolini:2023kxp} for a more complete discussion.}, which can either be isolated points (called nuts) or two-dimensional surfaces (called bolts) \cite{BenettiGenolini:2023kxp}. An essential feature of the fixed point sets is that they are in one-to-one correspondence with the the locations where the Killing spinor is chiral (without vanishing) $\eta = \pm \gamma_{(4)} \eta$, with $\gamma_{(4)}$ the four-dimensional chirality operator.

Referring back to the saddle point expansion in \eqref{Eq: QG saddle expansion}, we would like to reinterpret the fixed point result of \cite{BenettiGenolini:2019jdz,BenettiGenolini:2023kxp} in terms of M2-branes. Presumably, the two derivative supergravity action arises from the (suitably regularized) one-loop quantisation of point-like M2-branes. The position of the branes constitutes eleven bosonic zero-modes which are integrated over. The fixed point formula tells us that instead of evaluating the full integral over the moduli space of zero-modes, one merely has to evaluate their contribution at the fixed points of $\xi$. We would like to emphasise that this localisation of zero-modes to the fixed point set of $\xi$ is established at the one-loop level (i.e. two-derivative supergravity) and may be corrected at higher loops. In \cite{Genolini:2021urf} it was observed that a modified fixed point formula still holds for the higher derivative action obtained in \cite{Bobev:2020egg}. In order to prove that localisation happens to all order we should follow the standard route of localisation and add 
a chosen $Q$-exact functional to the action and take the $\hbar\to0$ limit. This task seems a tall order which is currently out of reach. In the remainder of this paper we will provide further evidence for localisation in M-theory by studying M2-branes with finite classical action and their one-loop quantisation. 

%%%%%%%%%%%%%%%%%%%%%
\section{Localisation in M-theory}\label{Sec: localisation in M-theory}
%%%%%%%%%%%%%%%%%%%%%
Let us now consider probe M2-branes in the eleven-dimensional geometry \eqref{Eq: 11d metric and three-form}. Let us furthermore demand that the M2-brane is BPS with respect to (some of) the supercharges defined by \eqref{11dspinor}. 
For a BPS brane configuration, the supersymmetry variation of the fermions should be cancelled by the local $\kappa$-symmetry transformation, which can be achieved iff \cite{Bergshoeff:1987dh}
\begin{equation}
	(1 - \rmi q \Gamma_{\text{M2}}) \epsilon = 0 \,,
\end{equation}
where $q=\pm1$ is the charge of the brane and $\Gamma_{\text{M2}} = \frac{1}{3!} \varepsilon^{abc}\Gamma_{abc}$ such that $\Gamma_\text{M2}^2=-1$.

Many different solutions exist for this equation, but since we are interested in M2-branes wrapping closed 3-manifolds with finite classical action, we will restrict to submanifold of $S^7$. The M2-brane projector can be rewritten in terms of the lower-dimensional gamma-matrices as
\begin{equation}
	\Gamma_{\text{M2}} = \gamma_{(4)}\gamma_{\text{M2}}\,,
\end{equation}
where $\gamma_{\text{M2}}$ is the product of the  pull-back of seven-dimensional gamma matrices to the worldvolume.
The BPS-condition 
can be solved separately for the four-dimensional spinor and the seven-dimensional spinor
\begin{equation}\label{separatedKappaEquation}
	\rmi \gamma_{\text{M2}} \chi^{\pm} = s\chi^{\pm} \,,\quad  \gamma_{(4)} \eta = sq \eta\,,
\end{equation}
where $s=\pm 1$.
Importantly, we find for the brane to be supersymmetric it must localise to the points where the four-dimensional spinor is chiral, i.e. at the fixed points of the Killing vector $\xi$. 
We thus find that the instantonic M2-branes localise on the nuts and bolts in four dimensions, just as the supergravity observables did!

The remaining task is to classify the solutions to the first equation in \eqref{separatedKappaEquation}. We focus on dominant solutions that have the smallest on-shell action. There are two possible configurations that are not related by symmetries of the background that must be considered. Roughly they are in one-to-one correspondence with the two projectors that appear in \eqref{the7dspinor}. For one solution the M2-brane wraps $(\theta_2,\phi_2,\varphi)$ and $\theta=0$, while for the other the brane wraps $(\theta,y,\varphi)$. In both cases the M2 worldvolume is a round Lens space $S^3/\Z_k$ with radius $2L$. Naively a third solution exists where the M2-brane wraps  $(\theta_1,\phi_1,\varphi)$ and $\theta=\pi/2$, but this is related to the first configuration by a $\Z_2$ symmetry of the background and should not be counted. Both inequivalent configurations have $s=1$ and have the same on-shell action
\begin{equation}
S_\text{cl} = \f{1}{(2\pi)^2 \ell_p^3}\int \text{vol}_3 = 2\pi \sqrt{{2N}/{k}}\,,
\end{equation}
where $\text{vol}_3 $ is the volume-form on the brane and we have used the relation between the gravity and field theory parameters $(L/\ell_{p})^6 = N k \pi^2/2$.

%%%%%%%%%%%%%%%%%%%%%
\section{Quantised M2-branes}\label{Sec: Quantisation of M2-branes}
%%%%%%%%%%%%%%%%%%%%%

We now turn to the one-loop partition function of the M2-branes. This involves quantizing the fluctuations of the brane around its classical configuration. The degrees of freedom consist of 4 complex scalar fields and 8 fermions living on the brane worldvolume. Since to this order the action is quadratic, the partition function is given by the determinant of kinetic operators
\begin{equation}\label{1loopdef}
Z_\text{1-loop} = Z_\text{zero-modes}\frac{\prod_f  \sqrt{\det'\cal D}}{\prod_b  \sqrt{\det'\cal K}}\,,
\end{equation}
where we have separated out the zero-modes from the determinants and collected into $ Z_\text{zero-modes}$. 

The kinetic operators ${\cal D}$ and ${\cal K}$ are computed by expanding the M2-brane action \cite{Bergshoeff:1987cm} to quadratic order. 
 The general formula for the bosonic operators in terms of background quantities is presented in \cite{Gautason:2024nru,Astesiano:2024sgi} and fermions can be treated using expressions in \cite{Harvey:1999as,Beccaria:2023sph,Astesiano:2024sgi}. Using these, the derivation of the operators is straightforward and hence we will not go through the explicit computation here. A detailed derivation will be presented elsewhere \cite{futurepaper}.  

The two supersymmetric M2-brane configurations discussed above lead to the same spectrum of fluctuation with identical kinetic operators. Their two partition functions therefore completely agree. For this reason we will focus on the first of the two configurations for which the worldvolume metric on $S^3/\Z_k$ takes the form
\begin{equation}
\dd s_\text{M2}^2 = L^2 \Big(\dd\theta_2^2 + \sin\theta_2^2 \,\dd\phi_2^2 +  (\dd \varphi -\cos\theta_2\,\dd\phi_2)^2\Big)\,.
\end{equation}
For the present case, the M2-brane worldvolume is particularly symmetric and therefore the kinetic operators are restricted to the form
\begin{equation}
{\cal K}=-D ^2 +  M^2 \,,\quad 
{\cal D}= i\slashed{D}  + M \,,
\end{equation}
where $M$ denotes the mass of the respective field and $D_a$ is a gauge covariant derivative which depends on a background gauge fields that arises due to the curvature and four-form field strength present in the background geometry. In this case two gauge fields arise, and the charge of each field with respect to these two gauge fields is determined, like the mass, by expanding the M2-brane action (see \cite{futurepaper}). The two gauge fields are given by
\begin{equation}
{\cal A}_{1} = \f12\big(\dd\varphi-\cos\theta_2\,\dd\phi_2\big)\,,\quad {\cal A}_2 = -\f12\cos\theta_2\,\dd\phi_2\,,
\end{equation}
and the covariant derivative is $D = \nabla - iQ_1{\cal A}_{1}- iQ_2{\cal A}_{2}$. % where $Q_{1,2}$ are the charges.

When quantizing the fields, it is efficient to perform a mode expansion along $\varphi$ for which the charge $Q_1$ shifts the mode numbers. The gauge field ${\cal A}_2$ on the other hand acts as a unit flux monopole field on the $S^2$ parametrized by $(\theta_2,\phi_2)$.

	\begin{table}[h!]
	\begin{center}
	{\renewcommand{\arraystretch}{1.1}
	\begin{tabular}{@{\extracolsep{10 pt}}llccc}
	\toprule
	Field & d.o.f. & $ Q_1 $   & $Q_2 $      & $ML$  \\\midrule
	Scalars &$4$ & $f$ & $0$ & $if/2$ \\
	& $4$ & $ 1$ & $ 1$ & $ i\sqrt{3}/2 $ \\\midrule
	Fermions & 	$4$ & $0$ & $0$ & $3q/4$  \\
	&$2$ & $ 1 + f  $ & $1$ & $-q(3/4 - f/2)$ \\
	&$2$ & $ 1 - f $ & $1$ & $-q(3/4+f/2)$\\
	\bottomrule
	\end{tabular}}
	
	\caption{Spectrum of scalar and fermion fluctuations. 
	}\label{Tab: 3d spectrum}
	\end{center}
	\end{table}
We are now in position to list the spectrum of fluctuations for the four scalar and eight fermionic fields. It turns out that the spectrum is completely determined by the self-dual or anti-self-dual component of the four-dimensional field strength evaluated on the brane. More precisely let us define
\begin{equation}\label{fdef}
f = \f12 |F + q\star_4 F|\,,
\end{equation} 
where the norm of a four-dimensional tensor is defined as $|T| = \frac12 \sqrt{T^{\mu\nu}T_{\mu\nu}}$ and this expression should be understood as being evaluated on the M2-brane worldvolume which we recall is located where the four-dimensional spinor satisfies $\gamma_{(4)} \eta = q\eta$. As can be observed from Table \ref{Tab: 3d spectrum}, the entire spectrum is controlled by $f$. 
The limit $f\to 0$, which is relevant for the AdS$_4$ background, recovers the spectrum computed in \cite{Beccaria:2023ujc}.

\paragraph{Zero-modes.}  For $k > 2$, the zero-mode sector is independent of $f$, and as such identical to the zero-modes studied in \cite{Gautason:2023igo}. Namely, there are twelve bosonic and twelve fermionic zero-modes.  %which were lifted in \cite{Gautason:2023igo} through a supersymmetric deformation, out of which it was determined that their contribution to the path integral equals $Z_{\text{zero-modes}} = 2$.
In \cite{Gautason:2023igo} a supersymmetric deformation was employed to lift the zero-modes and by taking the deformation to zero, their contribution was determined to equal $Z_{\text{zero-modes}} = 2$.
Our analysis provides an alternative perspective on this result: imposing supersymmetry with respect to the spinor \eqref{11dspinor} leads to two equally contributing M2-brane configurations, explaining the factor 2.

\paragraph{The one-loop determinants.} In order to evaluate the determinants in \eqref{1loopdef}, we follow the same strategy as in \cite{Beccaria:2023ujc} which was to expand in fourier modes along $\varphi$ and quantize the tower of 2D modes using the results of Section 2 in \cite{Gautason:2023igo}. Although worldvolume supersymmetry is not manifest in the Green-Schwarz formalism, it does seem to reveal itself in the spectrum by the fact that there is an enormous cancellations between fermionic and bosonic contributions. The contribution of the remaining unpaired modes can be expressed in terms of the following infinite products
\footnote{Note that $s(z)$ is nothing more than the standard 3d $\mathcal N=2$ chiral multiplet contribution to supersymmetric $S^3$ partition function.}
\begin{equation}
	s(z) = \prod\limits_{n=1}^{\infty} \left( \frac{n+z}{n-z} \right)^n\,,\quad t(z) = \prod\limits_{n=1}^{\infty} \frac{k^2}{4}(n^2-z^2)\,,
\end{equation}
which can be evaluated using $\zeta$-function regularization 
\begin{equation}
\begin{aligned}
	s(z) =&\, \rme^{\frac{\rmi  \text{Li}_2(\rme^{2\pi\rmi z})}{2\pi}- \frac{\rmi \pi}{12}  - z\log\left[ 1-\rme^{2\pi\rmi z} \right]+\frac{\rmi\pi z^2}{2}}\,,\\
	t(z)=&\, \frac{4}{k}\frac{\sin\pi z}{z}\,.
\end{aligned}
\end{equation}
We are now in position to write the full one-loop partition function of instantonic M2-branes by combining the zero-modes and non-trivial determinants and summing over the fixed point set in four dimensions
\begin{equation}\label{Eq: final answer}
Z_{\text{M2}} = 2 \sum\limits_{\substack{\text{fixed}\\\text{points}}} \frac{s(\tfrac{2}{k})^{2k}s(x_+)^{-k}s(x_-)^{-k}}{t(x_+)t(x_-)}\rme^{-2\pi \sqrt{2N/k}}\,,
\end{equation}
where we have introduced $x_\pm = (2/k)(1\pm f)$ and the sum runs over the fixed points of the Killing vector $\xi$. We have not distinguished between isolated fixed points and fixed surfaces but the latter should be integrated over with a suitable measure which we will discuss momentarily. We would like to emphasise that this result holds for any supersymmetric background of minimal four-dimensional $\mathcal N=2$ gauged supergravity. For a given four-dimensional solution and its fixed point set, the chirality constraint determines the M2-brane charge $q$ which in turn feeds into the definition of $f$ and $x_\pm$ in \eqref{fdef}.

%%%%%%%%%%%%%%%%%%%%%
\section{Examples}
%%%%%%%%%%%%%%%%%%%%%
Let us now apply our formula \eqref{Eq: final answer} to a few well known solutions of four-dimensional supergravity. We will be brief but a more complete treatment will appear in \cite{futurepaper}.

\paragraph{(I) AdS with a round $S^3$ boundary.} The R-symmetry Killing vector has an isolated fixed point at the centre of AdS, where both charges for the M2-brane are supersymmetric. For both charges $f=0$ such that 
\begin{equation}\label{roundS3}
Z_{\text{M2}} = \frac{1}{\sin^2(2\pi/k)}\rme^{-2\pi \sqrt{2N/k}}\,,
\end{equation}
recovering the result of \cite{Beccaria:2023ujc}. This answer matches the leading non-perturbative corrections to the ABJM partition function on the round sphere \cite{Hatsuda:2012dt}.

\paragraph{(II) AdS with a squashed $S^3_b$ boundary, preserving a $\text{U}(1)^2$ isometry \cite{Martelli:2011fu}.}  %, with metric

The fixed point of the Killing vector is at the centre of AdS and the spinor has a definite chirality, allowing only a single charge for the M2-brane to be supersymmetric. 
The Yang-Mills field is non-trivial, such that $f= \frac{b^2 - 1}{b^2 + 1}$ and
\begin{equation}\label{squashedS3}
Z_{\text{M2}} = 2\frac{s(\tfrac{2}{k})^{2k}s(\tfrac{4b^2/k}{b^2+1})^{-k}s(\tfrac{4/k}{b^2+1})^{-k}}{t(\tfrac{4b^2/k}{b^2+1})t(\tfrac{4/k}{b^2+1})}\rme^{-2\pi \sqrt{2N/k}}\,.
\end{equation}

\paragraph{(III) Supersymmetric extremal AdS-Kerr-Newman black hole \cite{Carter:1968ks,Kostelecky:1995ei,Caldarelli:1998hg}.} There are two fixed points, at the north- and south-pole of the horizon, where the Killing spinor has a definite chirality and $f = \frac{\omega-1}{\omega+1}$, with $\omega$ the angular chemical potential of the black holes, in the conventions of \cite{Cassani:2019mms}. The sum in \eqref{Eq: final answer} therefore reduces to a factor 2 and the M2-brane partition function can be read off. The dual partition function is related to the superconformal index of the ABJM theory \cite{Bhattacharya:2008zy,Bhattacharya:2008bja}.

\paragraph{(IV) Supersymmetric thermal AdS, i.e. Euclidean AdS in global coordinates with a non-contractible time circle.} In contrast to the black hole background, the Killing vector does not have fixed points, and thus there are no supersymmetric (instantonic) M2-brane configurations. This is consistent with recent results which showed that the dual partition function, which is related to the 1/2-BPS index of the ABJM theory, can be written as a giant graviton expansion of M5-branes \cite{Arai:2020uwd,Gaiotto:2021xce,Beccaria:2023cuo}. 

\paragraph{(V) Euclidean dyonic black hole with Riemann surface horizons of genus $\mathfrak g>1$ \cite{Romans:1991nq}.} The dual partition function is related to the topologically twisted index (TTI) \cite{Benini:2015noa, Benini:2016hjo,Closset:2016arn}, which was argued to equal the superconformal index and the $S^3_b$ partition functions in the ``Cardy-limits'' $\omega\rightarrow 0$ and $b\rightarrow \infty$ \cite{Choi:2019dfu,Bobev:2024mqw}, up to a pre-factor $(1-\mathfrak{g})$ \cite{Benini:2016hjo,Bobev:2022eus}. This is consistent with the fact that $f = 1$ for all three associated backgrounds in these limits. The fixed point set in this case is the black hole horizon, where the spinor has a definite chirality, and the fixed point sum in \eqref{Eq: final answer} should be treated as an integral. We determine the integration measure from the aforementioned relations between the holographically dual partition functions, such that the integral evaluates to  $(1-\mathfrak g)$ \footnote{This is in line with the localisation in the moduli space of the point-like brane, i.e. supergravity \cite{BenettiGenolini:2023kxp}, and with the absence of additional zero-modes implying that no factors of the target space length scale should be introduced when evaluating the integral.}.
\section{SUSY Localisation}

A fascinating feature of the result \eqref{roundS3} is that even though it comes from a one-loop quantisation, it matches the leading non-perturbative correction to the field theory grand canonical $S^3$ partition function \cite{Marino:2011eh}
\begin{equation}
	\rme^{J(\mu,k)} =  \sum_{N=0}^\infty Z_{S^3}(N,k)\rme^{\mu N}\,.
\end{equation}
A possible explanation is that in string/M-theory we compute the partition function for a fixed background, and thus for fixed length scale $L/\ell_p$. The canonical ABJM partition function is instead computed for a fixed rank $N$ and consequently for a holographic comparison one should perform a Legendre transform interchanging between keeping $\mu\sim(L/\ell_p)^3$ fixed and keeping the flux quantum $N$ fixed. 

The feature of one-loop exactness seems to persist more generally for non-perturbative corrections to the grand canonical potential $J(\mu,k)$ that can be mapped to (worldsheet) instantons in the gravitational theory. Namely, the coefficients of these exponentially suppressed terms are determined by a %Gopakumar-Vafa invariant on 
topological string on local $\mathbf{P}^1 \times \mathbf{P}^1$, and factors of $\sin 2\pi/k$ \cite{Hatsuda:2012dt}. As such they are independent of $N$ and consequently are determined in terms of a one-loop quantisation of instantonic M2-branes.  

%This remarkable fact suggests that the quantisation of  M2-branes in the AdS$_4\times S^7/\Z_k$ geometry admits some form of supersymmetric localisation. 

The fact that the one-loop result in \eqref{roundS3} matches the non-perturbative correction in the field theory grand canonical partition function could be an artefact of the large amount of supersymmetry, which is not present for the more general setup studied here. Currently there are only a handful of ABJM observables which have been determined non-perturbatively. 
Notably, for the squashed sphere partition function with $b=\sqrt{3}$ of the $k=1$ ABJM theory, results of \cite{Hatsuda:2016uqa} support the one-loop exactness of M2-branes we have discussed here. The reasons are similar as for the round $S^3$ partition function. Namely, there is a topological string description for the non-perturbative sector that dictates a simple $N$ dependence of the non-perturbative effects in the grand canonical ensemble. 
Recently, similar results were obtained for a mass-deformed ABJM on the squashed sphere 
\cite{Nosaka:2024gle,Kubo:2024qhq}.

%, again signalling the one-loop exactness of the quantum M2-branes in these backgrounds. 

In this letter we demonstrated that by picking a suitable supercharge the M2-brane is localised in target space eliminating many of the subtleties encountered in \cite{Gautason:2023igo,Beccaria:2023ujc}. 
%We have not demonstrated the one-loop exactness of the M2-brane partition function which presumably results out of localisation with respect to the chosen supercharge. 
By analysing a broad class of observables and performing explicit one-loop analysis in M-theory, we have uncovered a formula \eqref{Eq: final answer} that gives tantalizing hints at the presence of localisation within quantum gravity. Recall that the supersymmetric configurations we studied involve M2-branes wrapping a lens space $S^3/\Z_k$ embedded in a flux background. The curvature and flux of the environment leads to a deformation of the worldvolume theory in the form of non-trivial background fields. We may speculate that at low energies the dynamics of these M2-branes are governed by a similarly deformed version of the ABJM theory and hence the M2-brane partition function may be obtained as the ABJM partition function which can be computed using localisation. 
Our one-loop result in \eqref{Eq: final answer} features basic building blocks of 3D supersymmetric partition functions familiar from the localisation literature \cite{Kapustin:2009kz,Jafferis:2010un,Hama:2010av}.  
A point we aim to address in the future is to directly reproduce \eqref{Eq: final answer} as an ABJM partition function.
	
We would like to emphasise that the unreasonably simple answer in \eqref{Eq: final answer} is a consequence of quantising instantons directly in M-theory, instead of weakly coupled string theory. Indeed, in the latter this answer can be thought of as an infinite series over loop diagrams of worldsheet instantons, corresponding to the large $k$ expansion of \eqref{Eq: final answer}. It will be very interesting to study whether localisation is also applicable to the leading saddle in the quantum gravitational path integral \eqref{Eq: QG saddle expansion}. Our results suggest that this study might be under better control in the context of M-theory, than type II string theory, and thus motivates us to revisit the approach of Fradkin and Tseytlin \cite{Fradkin:1984pq,Fradkin:1985fq,Fradkin:1985ys} in the context of degenerate M2-branes, for example along the lines of \cite{Berman:2006vg}.

%%%%%%%%%%%%%%%%%
%\section*{Acknowledgements}
%%%%%%%%%%%%%%%%%
We are thankful to Nikolay Bobev, Junho Hong, Valentina Puletti, and Valentin Reys for initial collaboration and discussions. We are especially grateful to Nikolay Bobev, Sameer Murthy, and  Arkady Tseytlin for useful comments.  FFG is supported in part by the Icelandic Research Fund under grant 228952-053. JvM is supported by the STFC Consolidated Grant ST/X000575/1 and acknowledges the ERC-COG grant NP-QFT No. 864583.

\bibliography{LocalisingM2sbib.bib}

\end{document}